\documentclass{article}

\usepackage{graphicx}        % standard LaTeX graphics tool
\usepackage{multicol}        % used for the two-column index
\usepackage[bottom]{footmisc}% places footnotes at page bottom

\usepackage{authblk}
\title{How low-cost AI universal approximators reshape market efficiency}

\author[1]{Paolo Barucca}
\affil[1]{Department of Computer Science, University College London, WC1E 6BT London, United Kingdom}

\author[2]{Flaviano Morone}
\affil[2]{Center for Quantum Phenomena, Department of Physics,
New York University, New York, NY 10003 USA}

\makeindex          

\begin{document}

\maketitle

\abstract{The efficient market hypothesis (EMH) famously 
stated that prices fully reflect the information available 
to traders \cite{fama1970efficient}. 
This critically depends on the transfer of information into 
prices through trading strategies. 
Traders optimise their strategy with models of increasing 
complexity that identify the relationship between information 
and profitable trades more and more accurately. 
Under specific conditions, the increased availability of low-cost universal approximators, such as AI systems, should be 
naturally pushing towards more advanced trading strategies, 
potentially making it harder and harder for inefficient traders 
to profit. 
In this paper, we leverage on a generalised notion of market efficiency, based on the definition of an equilibrium price process, 
that allows us to distinguish different levels of model complexity through investors' beliefs, and trading strategies optimisation, 
and discuss the relationship between AI-powered trading and the time-evolution of market efficiency. 
Finally, we outline the need for and the challenge of describing out-of-equilibrium market dynamics in an adaptive multi-agent environment.}

\section{Introduction}
\label{sec:1}
According to many economists markets are the most efficient 
mechanism for collecting, aggregating, and condensing widely 
diverse information into a single estimate expressed by the 
price~\cite{black1986noise}. 
Markets do that with trading. In financial markets, traders 
collectively shape price 
dynamics by mapping their information into trades. 
If many traders agree that an asset is worth more than what it is selling for, they will try to buy it, thus collectively pushing the price up and eliminating the potential arbitrage. 
In doing so, traders' individual judgements, and the 
information those judgements are based on, become `priced in 
the market'. 
Thus, the process of seeking a profit 
ends up enhancing the efficiency of financial markets.
\\
The theory of market efficiency, \cite{fama1970efficient}, in its original strong formulation, hypothesizes that prices 
reflect all the available information at all times and that 
this should imply that price changes will be independent and 
identically distributed, and under specific conditions follow a random walk, \cite{bachelier1900theorie,samuelson2016proof}.
This original hypothesis did not explain what happens when new information is incorporated, as it implied an instantaneous transmission of new information into price changes. 
This information transmission is paradoxical: price changes only occur as a consequence of a trade, a trade that should be based on new information not reflected in the price until it occurs. Hence, informed trades should exist in order for the price to adjust to new information,\cite{grossman1980impossibility}. 
\\
Despite its paradoxical nature, the theory of efficient markets does capture a fundamental understanding of market functioning in finance, as it gives a framework to discuss the connection between the (im)possibility of traders to make a profit based on the available information and the properties of price time series. 
For example, as shown in \cite{delbaen1994general}, the absence of arbitrage opportunities should lead to the existence of an equivalent local martingale measure for the price process. 
\\
The different original formulations of the theory of market efficiency, \cite{fama1970efficient}, distinguish between discrete levels of information, i.e. public and private, but do not clarify what are the implications for a trader operating in mixed marketplaces where information is unequally distributed and traders use models with different levels of intelligence, i.e. with a different ability to extract signal from a given information set. 
Furthermore, the hypothesis is largely independent on the capability of traders to make new trades, 
due to endogenous constraints, such as dry powder, or on the market power and potential market impact of the trades of large investors, but also exogenous factors, such as high yield rates, and thus putting the {\it ability} to trade on an equal footing with the {\it willingness} to trade.
\\
In this paper, we consider recent advances in the understanding of market efficiency, such as \cite{timmermann2004efficient} and \cite{jarrow2012meaning}, which account for trading strategies, information asymmetry and model complexity, and use it to formulate a series of hypotheses on the expected evolution of (in)efficient markets in presence of a growing fraction of AI traders.
Previous works have discussed, also critically, the impact of the whole algorithmic trading and automation on market efficiency, \cite{yadav2015algorithmic}, and only recently scholars have directly addressed the question of the specific role of AI in shaping market efficiency \cite{marwala2017efficient,mancuso2022efficient}. 
Here, leveraging on a consolidated mathematical framework used to model market efficiency, we are able to discuss different levels of market efficiency and discuss why low-cost universal approximator, such as AI systems, constitute a potential novelty in the evolution of market efficiency. We hypothesize that the deployment of AI models which are able to train larger and larger classes of models at lower optimisation costs should force traders to follow higher standards of efficiency while minimizing the use of human-defined trading strategies. 
\\
Finally, we discuss the role of population dynamics in the emergence of market efficiency and the necessity of detailed agent-based models and game-theoretical analysis to understand the functioning of financial markets, especially in the presence of AI-human interactions. 

\section{Literature review} 
\label{sec:2}
Market efficiency is a fundamental concept in both finance 
and economics and has been investigated from many perspectives 
and with different levels of mathematical rigor. In this study, we mainly focus on the direct ramifications of the seminal work on the efficient market hypothesis by Fama \cite{fama1970efficient}, which focuses on the relationship between the information available to traders, their ability to make predictable profits, and how this is reflected in the statistical properties of financial returns. 
\\
Famously, the efficient market hypothesis has three formulations: the weak, semi-strong, and strong one. In the weak formulation, the statement is the following: no trader is able to make a predictable profit solely based on available public market information. Its semi-strong formulation reads: no trader is able to make a predictable profit based on all the available public information (market and beyond). Finally, its strong formulation states: no trader is able to make a predictable profit based on both public and private information, i.e. no predictably profitable trading strategy is possible in a strongly efficient market. In other words, \cite{fama1970efficient}, “The strong form tests of the efficient markets model are concerned with whether all available information is fully reflected in prices in the sense that no individual has higher expected trading profits than others because he has monopolistic access to some information."
This notion was initially, and still is at times, considered as an indication that neither technical nor fundamental analysis or even insider trading, can allow traders to achieve returns greater than those that could be obtained by a random buy and hold portfolio with the same level of risk, \cite{malkiel1989efficient}. 
\\
Market efficiency, as a blanket concept that hypothesizes collective forecasting consequences on price time series due to the strategies of traders, was expanded and adapted by \cite{timmermann2004efficient}. In this work, the authors introduced a model-specific definition of market efficiency, where price changes are independent and unpredictable conditioned on the forecasting model that the traders are using to map the available information into predictions: in an efficient market there exists an entire set of models for which the expectation of future discounted return is zero. Moreover, the authors introduce a notion of time locality in market efficiency, i.e. the possibility for markets to be temporarily inefficient and then become efficient, or the other way around. 
Indeed the evolutionary dynamics of the market participants and of their trading strategies reflects the adaptive nature of market efficiency, \cite{lo2005reconciling},  and poses the challenge of modelling out-of-equilibrium market dynamics, i.e. how financial markets approach efficiency, or fall out from it.
\\
In \cite{lim2011evolution}, the authors observe that the possibility of weak-form time-varying market efficiency has received increasing attention in the last decades, showing as evidence the time-varying structure of autocorrelations measured in \cite{ito2009measuring}.
Although accounting for model specification, the model of market efficiency by \cite{timmermann2004efficient} is effectively limited by the need to introduce an arbitrary pricing kernel ({\it viz.} pricing model) to actually conduct a forecasting test over the returns. 
The need to assume a specific asset pricing model, the so-called ``joint hypothesis problem'', was tackled by the seminal paper, \cite{jarrow2012meaning}, where the authors combine the definition of an economy, consisting of a set of investors trading on a market, and define market efficiency as the ability of the financial market price to converge towards an equilibrium price process, thus avoiding the need to impose an asset pricing model.
\\
\\
Real markets cannot be regarded as such efficient aggregators of information that no trader should 
ever be so naive as to hope to beat them. Traders make mistakes, markets are not perfect, and cannot be regarded as strongly efficient, as noted early in the literature, \cite{gordon1985efficient}, and stated Fama himself, "We would not, of course, expect this model to be an exact description of reality, and indeed, the preceding discussions have already indicated the existence of contradictory evidence.” 
However, experience reveals that even if markets are less efficient than the strong form of the 
hypothesis supposes, it is still very difficult to consistently beat them, which is why so few traders can allegedly do it. 
Therefore, we can consider these definitions as ideal scenarios serving as a framework for developing tests for quantifying real market efficiency. 
\\
The notion of market efficiency as prices fully reflecting information can be interpreted in two main ways: (i) statistical unpredictability of the price time series and (ii) prices ought to be adherent to a fundamental value that the available information should entail \cite{black1986noise}. 
\\
These two interpretations, the statistical and the fundamental one, differ substantially in the ways they can be tested: whilst predictability of price time series can be formulated in terms of expectation over future profits, potentially including discount factors, dividends and pricing kernels \cite{walter2006martingales, timmermann2004efficient}, adherence to a fundamental value supposes the ability to build a structural model for the price formation based on the information available to traders. 
%For example, stock prices do not always reflect the  true value of companies, even though all the information is publicly  disclosed. Nevertheless, a trader should study this information thoroughly by really understanding the business model of the company, its capital  and management in order to develop a well-calibrated estimate of its real value and, in turn, make a profitable investment. 
%Furthermore, a structural model that turns information into  price estimated on one company will work poorly, if at all, on another.
\\
Measuring efficiency based on the discrepancy between asset prices and their fundamental values is difficult for one main reason: even if there were a structural model for the value of an asset, it would need to take into account the actual market where the asset is traded and the prices that traders would be willing to pay for it, hence it would need to model the market itself. 
Unsurprisingly, tests for market efficiency have focused on the statistical interpretation of the efficient market hypothesis, mainly testing the ability of factor models including available information to predict price returns, \cite{fama1969adjustment, fama1993common}. 
This major limitation was addressed in \cite{jarrow2012meaning}, where the authors introduce an economy associated with the market and define market efficiency as the correspondence between the security market price process and the equilibrium price process, \cite{duffie1986stochastic}, of a commodity economy. The authors avoid the ``joint hypothesis problem'': the necessity of testing market efficiency within a given equilibrium model. Their definition reconciles the statistical and fundamental interpretation of market efficiency by showing the correspondence between the notion of no-arbitrage, \cite{delbaen1994general}, in the security market, and the existence of an equilibrium price process for the underlying commodity market.
We will mainly use the formalism and equilibrium notions introduced in \cite{jarrow2012meaning} to discuss the out-of-equilibrium evolution of market (in)efficiency. 
\\
\\
\section{Market efficiency, investors' beliefs, and trading strategies}
The efficient market hypothesis models efficiency in terms of the statistical unpredictability of the return time series that derives from the necessity of the absence of arbitrage opportunities. It starts from an idealised view of financial markets where traders have access to the same information set, \cite{fama1970efficient}, and from that it hypothesizes that markets with informed traders should display randomly fluctuating returns with no arbitrage opportunities, or, more generally, efficiency should imply the existence of an equivalent martingale measure for the price process.
\\
The information set is a cumulative collection of time-structured data, including traditional macroeconomic time series, financial indicators, alternative data, high-resolution market data, fundamental data about individual stocks, firms, sectors, and more. In \cite{jarrow2012meaning} the information available to traders is expressed as a filtration over a probability space. 
The investors' beliefs in \cite{jarrow2012meaning} represent the level of model capacity the investors have achieved, i.e. they constitute a form of market intelligence. In this definition of market efficiency, investors' beliefs are crucial as they need to be equivalent to the measure of the probability space.
In \cite{jarrow2012meaning} the probability measure defines the price process, which is defined as efficient if it corresponds to the equilibrium price process of a security and commodity market. 
In this setting, model complexity is embedded in the probability measure and market efficiency implies the existence of an equivalent martingale measure. 
Both these generalised definitions allows us to consider market (in)efficiency not only at a global market level but at a trader's level. In particular, in \cite{jarrow2012meaning}, the economy associated to the market consists of a finite set of investors each provided with their beliefs, information, and utility functions. Beliefs, information, and utility functions all contribute to shape the individual investor's problem and consequently their optimal consumption and trading strategy. In this framework, in presence of market efficiency, all the individual optimal trading strategies are maximal, as demonstrated in \cite{jarrow2012meaning}, and, notably, are dominated by a simple market portfolio (and by each security holding). Given any market, to disprove efficiency one simply needs to find an arbitrage opportunity or a trading strategy dominating the market portfolio.
Arbitrage opportunities can emerge when the assumptions that lead to a market equilibrium are not met and investors fail to identify an optimal solution to the investor's problem, so that their consumption choices and the trading strategies are not optimal, and the market prices do not necessarily provide an equilibrium price process for the economy. 

\subsection{From efficient markets to efficient traders}
The efficient market in \cite{jarrow2012meaning} is associated with an equilibrium price process which embodies the structural model that should map all the available information into prices; all traders have access to the available information and their beliefs are equivalent to the price probability measure.
In real markets, traders differ based both on the information they can gather, their beliefs' and on the trading strategies they can build.
Formally, we have two separate objects: the complete filtered probability space of the market, and the investors' beliefs with their information. 
\\ 
We then proceed to classify different investors in the market based on their relative ability to detect signal in the market, i.e. to align their beliefs to the probability measure of the information set. In practice, this is an inference problem for traders who need to update their beliefs based on the observed information and price processes.
\\
Let us assume a market where the set of available information is the same for all traders, and it coincides with the set of information of the market itself. 
We can now consider two main cases: 
\begin{itemize}
    \item traders who are unable to align their beliefs with the probability measure of the process, and to find an optimal solution to the investor's problem (inefficient traders), 
    \item traders who are able to align their beliefs with the probability measure of the process and find an optimal solution to their investor's problem (efficient traders).
\end{itemize}
In this scenario, there is no guarantee for market efficiency, and as a consequence the no-dominance does not apply, i.e. there could be profitable admissible strategies  that outperform the market portfolio, and we expect the efficient traders to be the ones to identify them. 
\\
Efficient traders need both to identify the right probability measure and the optimal solution to the investor's problem. 
This comes at a cost, that we name optimization cost, defined as the cost needed to discover the probability measure, and, even if the true probability measure is available to efficient traders, they still may not be able to identify the optimal strategies. 
If only some traders succeed, then the market is inefficient, there exists a non-empty set of traders who can make predictably profitable trades, i.e. the market can exhibit arbitrage opportunities. 
In inefficient markets, efficient traders can beat the collective intelligence of the market. For a simple isolated small slow human investor using limited information, i.e. an uninformed simple trader, in an inefficient market, it should be almost impossible to do better than efficient traders. 
In inefficient markets competition is key and no trader can simply assume that a simple market portfolio will not be outperformed (on average) by better strategies. 
The emergence of market efficiency is positive especially for uninformed traders, i.e., when market efficiency applies, uninformed traders can simply invest in the market portfolio and forget about complex tasks such as aligning their beliefs and identifying complex (and expensive) trading strategies. 
This is a paradoxical perspective provided by market efficiency: it eliminates the out-of-equilibrium dynamic competition that is the very reason market efficiency should emerge. 

\subsection{Co-evolution of price and beliefs}
In real (in)efficient markets, we still expect arbitrage opportunities to disappear, i.e. inefficient markets should approach efficiency: we expect the distance between the equilibrium price process and the realised price process to remain small. 
As discussed in \cite{timmermann2004efficient}, ”An efficient market is thus a market in which predictability of asset returns, after adjusting for time-varying risk-premia and transaction costs, can still exist but only ‘locally in time’ in the sense that once predictable patterns are discovered by a wide group of investors, they will rapidly disappear through these investors’ transactions.” In other words, predictable patterns, when they exist, they tend to self- destruct after a certain period of time. Arguably, this is how the collective intelligence of markets should emerge: the more traders expand their information sets and improve their capacity of estimating beliefs, the more the set of probability measure should evolve and push the equilibrium price process to higher complexity levels.
\\
The presence of an arbitrage opportunity could be associated to information asymmetry and to the ability of some investors to better identify optimal trading strategies thanks to their better information gathering and belief estimation.
This fundamental mechanism creates a pressure for the time evolution of the equilibrium price process in efficient markets: the market price cannot be indifferent to investors' beliefs even %way 
when they are potentially far from the market probability measure as this would lead to the persistence of arbitrage opportunities.
This could be the case if investors were systematically unable to gather relevant information for the price process or to approximate the probability measure, due to modeling limitations. In this setting, for the market to be efficient it would need to temporarily co-evolve with the traders' beliefs, i.e. the probability measure would need to be aligned with the investors' beliefs and not the other way around, at least until traders were able to update their information and improve their modeling capacity. 
In particular, here we argue that the probability measure complexity should co-evolve with the complexity of the investors' beliefs in the market. 
In an evolving close-to-efficiency inefficient market, the probability measure should evolve over time, so that its complexity remains close but above the average complexity of the beliefs in the market. 
\\
So far we have considered traders with different beliefs' but sharing the same information.
In general, as originally discussed in \cite{fama1970efficient}, we can distinguish different information sets, for example market information on price time series, public information on companies, and private information on business strategies and more. 
The acquisition and integration of information in the models available to traders and in equilibrium price process are a fundamental part of the market evolution. 
In \cite{jarrow2012meaning}, the cases of information reduction and information expansion are considered: information reduction, namely the reduction of the filtration in the filtered probability space, is found to preserve market efficiency whilst information expansion is generally not. 
\\
The probability measure may be based on a large information set and traders may have access to limited information sets that hinder their ability to design profitable strategies. 
Conversely, the acquisition of new information by traders could lead to the identification of new equilibrium price processes depending on larger information sets.
The set of data that are now available about companies and their time resolution has significantly increased and, at the same time, this requires an increase in the models' ability to incorporate this information and translate that into optimal predictions. 
\\
The cost of collecting and processing this increasing amount of data pushes traders to improve their ability of extracting signal from them, i.e. a large investment on data collection needs to be coupled with an investment in the model complexity, at least up to a certain point, to make sure that all the trading signal that is present in the data is identified and exploited. 
This implies that the investors' beliefs and the equilibrium price process should co-evolve with the set of available information. More information requires more complex models, able to extract new signals from the new information; conversely the acquisition of new information pushes traders to improve their beliefs', %models, 
which as a consequence should push the equilibrium price process to higher complexity levels. 

\section{Low-cost universal approximators as drivers of market efficiency}
The time evolution of market efficiency has been investigated mainly through the study of the existence of regime changes and structural changes in autoregressive models, \cite{lim2011evolution}. 
Linear models have the advantage of efficient and reliable solvers to identify the optimal models, but only consider linear combinations of factors to determine the price. Universal approximators have been known and studied in functional analysis for a long time, \cite{stone1948generalized}, and recently multilayer perceptrons, \cite{hornik1989multilayer,barron1993universal}, and their generalisations, have been recognised and widely utilised as such. 
In particular, in \cite{hornik1989multilayer}, the authors demonstrated that multilayer feedforward networks are capable of approximating any Borel-measurable function from one finite dimensional space to another to any desired degree of accuracy, provided sufficiently many hidden units are available, and multiple other studies have extended and generalised this result, \cite{cybenko1989approximation,barron1993universal}. 
\\
The ability of approximating a function with a class of other basis functions is not a novel result, yet the limited generalisation errors and the efficiency of the training algorithms, yielding relatively low computational costs, for neural networks, \cite{rumelhart1986learning}, have notoriously established the use of these models in a wide range of applications. 
In finance, a growing number of studies is showing how AI algorithms can outperform traditional statistical models, e.g. \cite{fischer2018deep,makridakis2020m4}, and multiple independent surveys on financial firms have recognised that a vast majority of traders and investors, \cite{jung2019machine}, will be relying more and more on AI models for their financial decisions. 
\\
The ability of AI models of acting as universal function approximators, also in presence of different modalities of data, such as text, time-series, images, makes them suitable models to adopt in the space of available models for traders. Despite the existence of other universal approximators, AI models are unique in terms of multimodality, capacity, and adaptability. Linear approximators are generally not multimodal, not easy to extend in terms of model capacity (unless they take the form of AI models themselves), and less flexible in terms of available architectures and optimisation schemes. 
Universal approximation theorems guarantee that, provided the convergence of the training algorithm, with sufficient data all informed AI traders will never under-perform other traders and, if existent, should always be able to identify and approximate the probability measure of the equilibrium price process.  
We argue that, under these hypotheses, a market of rational traders will be inclined to adopt AI models and, as discussed above, the equilibrium price process should be pressured to become AI-efficient, i.e. not even AI models should, eventually, be able to identify profitable strategies. We may refer to this as asymptotic-AI-efficiency.
%AI-efficiency singularity. 
At that stage, the competition in the market will focus on the rapid integration of new information, rather than on the model itself. 
Convergence and big data are not guaranteed, so models will keep changing in terms of data available and in terms of players in the market. 

\subsection{Trading in a world of AI-efficient traders}
The notion of economy-dependent market efficiency, \cite{jarrow2012meaning}, is extremely useful to provide a richer picture of economic equilibrium in financial markets, yet it leaves us with complex questions over markets' evolution: how does the space of available information evolve? And, how do beliefs and price process co-evolve? 
\\
Hardware and software design has led to an increased capability to harness information and to an increased ability to explore the space of statistical models for estimating beliefs and developing trading strategies. 
The specific way in which the price process varies over time depends on the market fundamentals, market design, market microstructure, e.g. the liquidity of the order book, and ultimately on the traders themselves. 
\\
In a realistic setting, traders share a certain fraction of information, as well as certain classes of models, nevertheless it is also likely that each trader will have a slightly -or significantly- different set of information and, at the same time, a different class of models that they are able to use. 
Hence, we should look at the changes in market (in)efficiency in two main directions: one is a longitudinal time dimension, i.e. information and beliefs change and evolve over time; and another one is cross-sectional, i.e. in a given market at a given time multiple traders can coexist who do not share the same information and the same beliefs. 
\\
The absence of profitable trades is hypothesized as a consequence of the market ability to identify perfect prices based on the information available to different traders, and this efficiency is enabled by the complexity of the equilibrium price process that is accounting for the collective intelligence of the market. 
\\
So far we have made the simplifying assumption that the equilibrium price process at a given time is a function of the information set which is independent from the beliefs and strategies of individual traders. In reality markets do display some form of co-dependence and reflexivity: the performance of a strategy will not only be based on its expected value but also on the simultaneous strategies of the other traders, e.g. the specific sequence in order placement can significantly affect the price that a trader can obtain with respect to others. The dynamic time-dependent mapping between information and optimal trades will depend on the 
simultaneous, or quasi-simultaneous, decisions of the other traders, that will partially move the optimal solution with each of their trades. 
This can be partially understood in a game-theoretical way. A well-known example is the following. Consider a game where players (traders) must guess a number between 0 and 100 and the winner is the player who guesses the number which is closest to $2/3$ of the average guess. The Nash equilibrium, i.e. the solution to this game, is $0$, so in theory it is impossible to beat the crowd (the market) if every player uses the optimal strategy. But real markets are not perfect as we said. In fact, this contest was actually run by the {\it Financial Times} in 1997 and the result was that, although many contestants were able to figure out the Nash equilibrium and guessed 0, they were wrong in thinking that everyone else would be as smart as they were, \cite{thaler2016behavioral}. 
In fact the average guess was 18.9 and so the winning guess was 13. 
This experiment thus shows that the performance of a strategy in a real-market does indeed depends on the simultaneous strategies of the other traders and so to beat the market a trader must model also all other traders' behavior. 
\\
In the framework we presented for modeling efficiency, we essentially have two rate of changes in complexity, the change in market complexity, that is how rapidly the complexity of the probability measure changes over time and how rapidly the complexity of the beliefs' models available to traders change. 
\\
Traders' directly affect market efficiency and they compete with the market price dynamics they actually contribute to define. Market efficiency is related to the models available to the traders who populate it. Let us, for instance, consider a market populated with traders with limited information and limited model capacity. It is likely that a new trader with more information and more model capacity could enter the market and easily find profitable strategies, i.e. the market would be inefficient. The precise mechanism by which the traders' ability is translated into the collective intelligence of the market is generally dependent on the details of the market design. 
\\
This complex adaptation makes financial markets different from a static and predictable computational task, such as classifying pictures or even solving combinatorial optimisation problems, and constitutes the very reason a market of investments or hedge funds exists: profitable trades are the consequence of a powerful prediction model and of a successful behavioral agent-based model \cite{bouchaud2018agent}. 
Traders are both responsible for the emergence of the collective intelligence of the markets whilst being in competition with it and, as agents, are effectively following a noisy learning algorithm trying to optimise their different objective functions in a non-stationary multi-player environment.
\\

\section{Conclusions}
In this paper, we discussed the notion of complexity in market efficiency and the importance of its out-of-equilibrum dynamics.  
Introducing different levels of complexity can allow market observers to describe the evolution of market (in)efficiency over time and make sense of the ability of traders to adapt to changing market conditions. 
More complex models are now available to traders to map the information into beliefs and beliefs into trading strategies, potentially leading to a form of market singularity, i.e. where all traders can effectively learn accurate prices instantaneously. 
We discussed how the existence of universal approximators and their increasing availability and decreasing computational costs could affect the evolution of market efficiency. 
On one hand, low-cost AI approximators could push the boundary of market efficiency to a highly competitive global AI-efficiency. 
On the other hand, models are limited by convergence, computational power, and by the complex modeling interactions between strategies, whilst we are observing the complexification of the economic dependencies, of the interconnectedness of countries, firms, and supply networks, the increase in the responsiveness of consumers' preferences, which can all contribute to hinder the predictability of the market performance of different stocks and companies, so that even more information and more complex models will be needed to reach a given market performance, and markets will complexify at a rate not even AI-traders will be able to adapt to.  

\bibliographystyle{apalike}
\bibliography{AIbib}

\end{document}